\documentclass[11pt, conference, letterpaper, twocolumn, twoside]{IEEEtran}

\usepackage{cite}
\usepackage{graphicx}
\usepackage{subfigure}
\usepackage{amsmath}
\usepackage{epsfig}
\usepackage{times}

\IEEEoverridecommandlockouts

\newtheorem{pavikt}{\em Theorem}

\newcommand{\argmin}{\operatornamewithlimits{argmin}}

\begin{document}

\title{Coding, Scheduling, and Cooperation in Wireless Sensor Networks}
\author{\authorblockN{{\Large Samar Agnihotri and Pavan Nuggehalli}}\\
\authorblockA{Centre for Electronics Design and Technology, Indian Institute of Science, Bangalore - 560012, India\\
Email: \{samar, pavan\}@cedt.iisc.ernet.in}%
}

\maketitle

\begin{abstract}
We consider a single-hop data gathering sensor cluster consisting of a set of sensor nodes that need to transmit data periodically to a base-station. We are interested in maximizing the lifetime of this network. Even though the setting of our problem is very simple, it turns out that the solution is far from easy. The complexity arises from several competing system-level opportunities that can be harnessed to reduce the energy consumed in radio transmission. First, sensor data in a cluster is spatially and temporally correlated. Recent advances in distributed source coding allow us to take advantage of these correlations to reduce the number of bits that need to be transmitted, with concomitant savings in energy. Second, it is also well known that channel coding can be used to reduce transmission energy by increasing transmission time. Finally, sensor nodes are cooperative, unlike nodes in an ad hoc network that are often modeled as competitive. This collaborative nature allows us to take full advantage of the first two opportunities for the purpose of maximizing cluster lifetime.

In this paper, we pose the problem of maximizing lifetime as a max-min optimization problem subject to the constraint of successful data collection and limited energy supply at each node. This turns out to be an extremely difficult optimization to solve. Consequently, we employ a notion of instantaneous decoding to shrink the problem space. We show that the computational complexity of our model is determined by the relationship between energy consumption and transmission rate as well as model assumptions about path loss and initial energy reserves. We provide some algorithms, heuristics and insights for several scenarios. In some situations, our problem admits a greedy solution while in others, the problem is shown to be $\cal{NP}$-hard. The chief contribution of the paper is to illustrate both the challenges and gains provided by source-channel coding and scheduling.
\end{abstract}

\section{Introduction}
\label{Intro}
Minimizing energy consumption is one of the primary technical challenges in sensor networking. Many sensor applications such as habitat monitoring and industrial instrumentation envisage scenarios in which a large number of sensor nodes, powered by tiny batteries, will be actively deployed for months and even years. In many instances, it may not be possible to replace these sensor nodes once they run out of energy because the sensor nodes could be inaccessible (for example, embedded in concrete structures to sense stress levels). Replacing dead batteries in a sensor network consisting of a large number of nodes may also not be economically feasible. There is a now a broad consensus that aggressive system level strategies impacting many layers of the protocol stack need to be devised to meet the lifetime requirement of extant and future sensor networks.

In this paper, we consider a single-hop sensor cluster. Nodes in the cluster periodically sample a field and transmit the data directly to a central location or base-station. We are interested in minimizing the energy spent by these nodes in transmitting, with the objective of maximizing cluster lifetime. Sensor nodes also spend energy in receiving data, sensing/actuating and computation. The energy spent in sensing/actuating represents a fixed cost that can be ignored. The energy cost of receiving data can be easily incorporated in our optimization model. We assume computation costs are negligible compared to radio communication costs. This is debatable assumption in dense networks; we intend to incorporate computation costs in future work.

We believe that our model is useful because many popular proposals recommend organizing a sensor network into clusters \cite{leach}. Here each cluster elects a cluster-head (which we call base-station). Nodes communicate only through their respective cluster-heads. Approaches that maximize cluster lifetime can be thought of as being complementary to network-wide approaches such as energy-efficient routing. Moreover, our model is applicable to scenarios where a roving base-station moves from one cluster to another, gathering data.

We define cluster lifetime as the time until the first node in the cluster runs out of energy. While this is a somewhat pessimistic definition, we argue that a cluster will consist of relatively few nodes. The failure of even one such node can have disastrous consequences on the cluster's performance (for example, coverage). This definition also has the benefit of being simple and popular \cite{Tassiulas, Radha}. Other definitions proposed for network lifetime such as mean expiration time and time until a certain fraction of nodes fail are not appealing from a cluster viewpoint.

Somewhat to our surprise, we find that analyzing the performance of this simple model is far from trivial. The complexity arises from several competing system-level opportunities to be harnessed to reduce the energy consumed in radio transmission. First, sensor data in a cluster is spatially and temporally correlated. In \cite{Slepian}, Slepian and Wolf show that it is possible to compress a set of correlated sources down to their joint entropy, without explicit communication between the sources. This surprising existential result shows that it is enough for the sources to know the probability distribution of data generated\footnote{Actually knowledge of conditional entropies suffices.}. Recent advances \cite{Pradhan} in distributed source coding allow us to take advantage of data correlation to reduce the number of bits that need to be transmitted, with concomitant savings in energy. Second, it is also well known that channel coding \cite{Prabhakar} can be used to reduce transmission energy by increasing transmission time. Finally, sensor nodes are cooperative, unlike nodes in an ad hoc network that are often modeled as competitive. This collaborative nature allow us to take full advantage of the first two opportunities for the purpose of maximizing cluster lifetime.

Motivated by our definition of cluster lifetime, we pose the problem of maximizing lifetime as a max-min optimization problem subject to the constraint of successful data collection and limited energy supply at each node. This turns out to be an extremely difficult optimization to solve. Consequently, we employ a notion of instantaneous decoding to shrink the problem space. We show that the computational complexity of our model is determined by the relationship between energy consumption and transmission rate as well as model assumptions about path loss and initial energy reserves. We provide some algorithms, heuristics, and insights for several scenarios. In some situations, our problem admits a greedy solution while in others, the problem is shown to be $\cal{NP}$-hard. The chief contribution of the paper is to illustrate both the challenges and gains provided by source-channel coding and scheduling.

There is much related work in this area. Energy conscious networking strategies have been proposed by many researchers mainly at the MAC \cite{S-MAC} and routing layer \cite{Tassiulas, Bhardwaj, Raghavendra, Meng, Bhaskar, Li}. Our study was motivated by previous research in \cite{Baek, Cristescu}, which explicitly incorporate aggregation costs in gathering sensor data. In \cite{Cristescu}, the authors consider the problem of correlated data gathering by a network with a sink node and a tree communication structure. Their goal is to minimize the total transmission (energy) cost of transporting information. The first part of \cite{Baek} considers a model similar to ours, namely, that of several correlated nodes transmitting directly to a base station. However, both \cite{Baek} and \cite{Cristescu} are interested in minimizing total energy expenditure, as opposed to maximizing network lifetime. In the latter case, the optimal solution is shown in both papers to be a greedy solution based on ordering sensors according to their distance (which reflects data aggregation cost) from the base station. However, we show that this solution is not optimal for maximizing network lifetime. An early version of our ideas appeared in \cite{wowmom05}. This paper generalizes the work presented in \cite{wowmom05} and provides proofs of some key conjectures there.

In section \ref{model}, we present our system model and describe our notion of instantaneous decoding. In section \ref{gc}, we consider a general channel model which allows us to consider the joint impact of cooperative nature of the sensor nodes and source and channel coding on system lifetime. We prove that the both, the static and dynamic scheduling problems for the general channel model, are $\cal{NP}$-hard and the optimal dynamic scheduling strategy does better than optimal static scheduling strategy, in general. We also provide the geometric interpretation of the optimal solutions and the solution search procedures. As a special case of this problem, in section \ref{srra}, we consider a scenario which allows us to neglect the impact of transmission time allocation. This is similar to the scenario considered in \cite{Baek} and \cite{Cristescu}. Here we provide some key insights into the nature of the optimal solutions for both, static and dynamic scheduling, derived in \cite{wowmom05}.

\section{System Model}
\label{model}
We consider a network of $N$ battery operated sensor nodes strewn uniformly in a coverage area. Time is divided into slots or rounds. In each slot, sensors take samples of the coverage area and transmit the information directly to the base station. We model the sensor readings at node $k$ by a discrete random process $X^n_k$ representing the sampled reading value at node $k$ in the $n^{\textrm{th}}$ time slot. We assume that sensor readings in any time slot are spatially correlated. We ignore temporal correlation by assuming that sensor readings in different time slots are independent. However, temporal correlation can easily be incorporated in our work for data sources satisfying the Asymptotic Equipartition Property (AEP). The entropy of $X^n_k$ is denoted by $h_k$.

Initially, sensor node $k,\; 1 \le k \le N$, has access to $E_k$ units of energy. The wireless channel between sensor $k$ and the base station is described by a path loss factor $d_k$, which captures various channel effects such as distance induced attenuation, shadowing, and multipath fading. For simplicity, we assume $d_k$'s to be constant. This is reasonable for static networks and also in the scenarios where the path loss parameter varies slowly and can be accurately tracked. The network operates in a time-division multiple access (TDMA) mode. In each slot, every sensor is allotted a certain amount of time during which it communicates its data to the base station.

The general problem is to find the optimal rate (the number of bits to transmit) and transmission times for each node, which maximize network lifetime. Both the rate and time allocation are constrained. The rate allocation should fall within the Slepian-Wolf achievable rate region and the sum of transmission times should be less than the period of a time-slot (which is taken to be unity). Finding the optimal rate allocation is a computationally challenging problem as the Slepian-Wolf achievable rate region for $N$ nodes is defined by $2^N-1$ constraints. We simplify the problem by insisting that decoding at the base-station be instantaneous in the sense that once a particular node has been polled, the data generated at that node is recovered at the base-station before the next node is polled. This reduces the rate allocation problem to finding the optimal scheduling order, albeit at some loss of optimality. This loss of optimality occurs because our problem formulation assumes the separation between source and channel coding and it is well-known, \cite{CoverElGamal}, that the source-channel separation does not hold for the multi-access source-channel coding problem and Slepian-Wolf coding followed by channel coding is not optimal for the joint source-channel coding problem. Also, in general, turning a multiple-access channel into an array of orthogonal channels by using a suitable MAC protocol (TDMA in our case) is well-known to be a suboptimal strategy, in the sense that the set of rates that are achievable with orthogonal access is strictly contained in the Ahlswede-Liao capacity region \cite{Cover}. However, despite this fundamental sub-optimality, we argue like \cite{Servetto, Kumar} that there are strong economic gains in the deployment of networks based on such technologies, due to the low complexity and cost of existing solutions, as well as available experience in the fabrication and operation of such systems.

Let $\Pi$ be the set of permutations of the set, $\{1,2,\ldots,N\}$. The polling schedule followed by the network in any time slot corresponds to a permutation, $\pi \in \Pi, |\Pi| = N!$. Let $\pi(k)$ denote the $k^{\textrm{th}}$ node to be scheduled. Instantaneous decoding implies that the amount of data to be transmitted by node $\pi(k)$ is the conditional entropy of the data source at node $\pi(k)$, given the data generated by all previously polled nodes. We denote the amount of information generated by node $\pi(k)$ by $h_{\pi(k)}$. Our aim is to find the scheduling strategy (scheduling order and transmission time allocation) that maximizes network lifetime.

\section{General Channel Scenario}
\label{gc}
In this section, we consider the general channel coding scenario where the transmission energy is the convex decreasing function of the transmission time. For example, by inverting Shannon's channel capacity formula for the AWGN channel, it is straight-forward to show that transmission energy is a strictly decreasing convex function of transmission time \cite{Prabhakar}. Other channel coding situations lead to a similar result. In such a scenario, we not only have to find the optimal scheduling order, but also the optimum transmission times for each node. We consider two kinds of schedules, namely, static and dynamic. In static scheduling, the nodes follow the same fixed scheduling order in all time slots until the network dies. Under dynamic scheduling, we allow nodes to collaborate further by allowing them to employ different schedules in different time slots. More specifically, it is \emph{offline} dynamic scheduling, where before the actual operation of network starts, the base-station has already computed the optimum set of schedules and the number of slots for which each schedule is used, rather than \emph{online} dynamic scheduling, where only at the beginning of every polling slot, the base-station computes \emph{on fly} the optimum schedule for that time-slot, based on its knowledge of the latest state of the network.

Let $f(h,x)d$ be the energy required to transmit $h$ bits of information in $x$ units of time with path loss factor $d$. So, we can interpret $f(h,x)$ as the energy required to transmit $h$ bits of information in $x$ units of time with unit path loss. Based on our discussion, we model the energy function $f(h,x)$ as follows.
\begin{enumerate}
\item $f(h,x)$ is a strictly \emph{decreasing} continuous positive convex function in $x$.
\item $\lim_{x \rightarrow 0} f(h,x) =  \infty$
\end{enumerate}

Unless stated otherwise, we assume $f(h, x)$ to be the one that is obtained by inverting Shannon's AWGN capacity formula \cite{Prabhakar}, that is:
\begin{equation}
\label{energy_fn}
f(h, x) = x(2^{\frac{h}{x}} - 1)
\end{equation}

\subsection{Static Scheduling}
\label{gc_ss}
In static scheduling, each permutation, $\pi \in \Pi$ corresponds to a TDMA schedule. Let $h_{\pi(k)} = H(X_{\pi(k)} | X_{\pi(k-1)}, \ldots, X_{\pi(1)})$ denote the number of information bits transmitted per slot by node $k$ under the schedule $\pi \in \Pi$. Let $t_{\pi(k)}$ be the corresponding transmission time alloted to node $k$ and $L_{\pi}$ be the lifetime achievable by the system under the schedule $\pi$. Note that lifetime is integer-valued but we will treat it as a real number. The optimal static schedule is the solution to the following optimization problem.
\begin{eqnarray}
\label{eqn:gc_ss}
\max_{\pi \in \Pi} L_{\pi} (& = & \min_{1 \le k \le N} \frac{E_k}{f(h_{\pi(k)}, t_{\pi(k)})d_k})\\
t_{\pi(k)} & \ge & 0 \nonumber \\
\sum^N_{k=1} t_{\pi(k)} & = & 1 \nonumber
\end{eqnarray}

However, using the ``Channel Aware'' algorithm proposed in \cite{wowmom05} for every schedule $\pi$, we have the maximum lifetime and the corresponding transmission times allocation vector $\{t_{\pi(k)}\}^{N}_{k=1}$. Further, for this transmission time allocation vector, all the sensor nodes achieve the same lifetime. So, the problem in \eqref{eqn:gc_ss} reduces to the following optimization problem:
\begin{eqnarray}
\label{eqn:gc_ss1}
\max_{\pi \in \Pi} L_{\pi} (& = & \frac{E_k}{f(h_{\pi(k)}, t_{\pi(k)})d_k})\\
t_{\pi(k)} & \ge & 0 \nonumber \\
\sum^N_{k=1} t_{\pi(k)} & = & 1 \nonumber
\end{eqnarray}

Before we analyze the solution of the optimal scheduling problem for the general case, it should be noted that given the cooperative nature of the sensor nodes, the nodes can collaborate with each other to a much greater extent by varying their transmission times. For example, nearby nodes can finish their transmissions sooner, allowing far away nodes more time to transmit in order to improve system lifetime. The structure of the problem in \eqref{eqn:gc_ss} or \eqref{eqn:gc_ss1} is such that its computational complexity depends on the sensor node data correlation structure as well as the energy function. The following three examples amply illustrate this.

\emph{Example 1:} Let us consider the following model for spatial correlation of the sensor data.

Let $X_i$ be the random variable representing the sampled sensor reading at node $i\in \{1, \ldots, N\}$ and $B(X_i)$ denote the number of bits that the node $i$ has to send to the base-station. Let us assume that each node $i$ has at most $n$ number of bits to send to the base-station, so $B(X_i) = n$. However, due to the spatial correlation among sensor readings, each sensor may send less than $n$ number of bits. Let us define a data-correlation model as follows.

Let $d_{ij}$ denote the distance between nodes $i$ and $j$. Let us define $B(X_i/X_j)$, the number of bits that the node $i$ has to send when the node $j$ has already sent its bits to the base-station, as follows:
\begin{equation}
\label{eqn:cor1}
B(X_i/X_j) = \left\{
                    \begin{array}{ll}
                     \lceil d_{ij} \rceil \mbox{ if } d_{ij} \le n\\
                     n \mbox{ if } d_{ij} > n
                    \end{array}
             \right.
\end{equation}

\begin{figure}[t]
\begin{center}
\includegraphics[angle=-90, width=8.85cm]{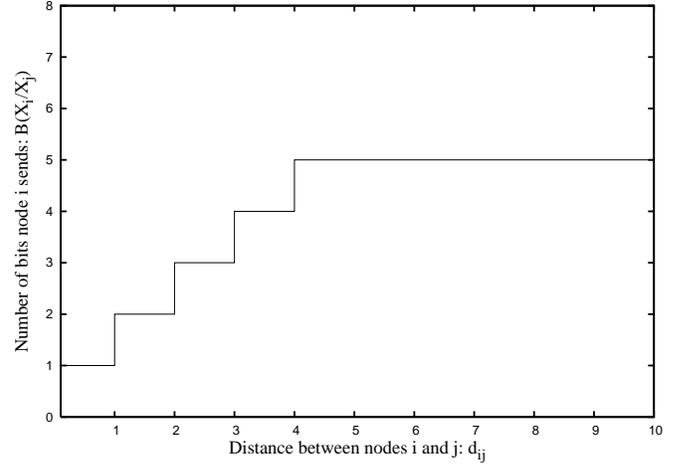}
\end{center}
\caption{Data Correlation Model for $n=5$: plot of $B(X_i/X_j)$ versus $d_{ij}$}
\label{fig1}
\end{figure}

Figure \ref{fig1} illustrates this for $n=5$. It should be noted that when $\lceil d_{ij} \rceil < n$, the data of the nodes $i$ and $j$ differ in at most $B(X_i/X_j)$ least significant bits. So, the node $i$ has to send, at most, $\lceil d_{ij} \rceil$ least significant bits of its $n$ bit data. Also note the following property of the correlation model:
\begin{equation}
\label{eqn:cor1prop}
B(X_i/X_1, \ldots, X_{i-1}) \le B(X_i/X_{i-1})
\end{equation}

However, the definition of the correlation model is not complete yet and we must give the expression for the number of bits transmitted by a node conditioned on more than one node already having transmitted their bits to the base-station. There are several ways in which this quantity can be defined. Presently, let us consider the following definition of the conditional information:
\begin{equation}
\label{eqn:gc_ss4}
B(X_i/X_1, \ldots, X_{i-1}) = \min_{1 \le j < i} B(X_i/X_j)
\end{equation}

\emph{Part 1:} Here let us assume that the ratio of energy of a node and path-loss between base-station and the node is equal for all the nodes. So, without the loss of generality, for every sensor $k$, we can put $E_k/d_k = 1, 1 \le k \le N$.

This assumption is only to simplify the solution, yet it is not such an unrealistic assumption when we consider the scenarios such as one where given the equal energies of all the nodes, the distance of the base-station from any node is much more than the distance between any two nodes. Also, when we have the roving base-station for the data gathering, this assumption holds good. So, using the energy consumption model of \eqref{energy_fn}, this makes the time to transmit depend only on the number of bits that a node has to send to the base-station, so a sensor polling schedule that results in larger value of $B(X_1, X_2, \ldots, X_N)$, will also result in the larger value of the sum of transmission times of all the nodes.

\begin{pavikt}
A greedy scheme that assigns information bits to the nodes according to \eqref{eqn:gc_ss4}, gives the solution for the static scheduling problem in \eqref{eqn:gc_ss1} in $\cal P$.
\end{pavikt}
\begin{proof}
Start with any node as the first node of the schedule, then choose the next node in the schedule to be that node that minimizes the conditional number of bits. However, given the definition of correlation model in \eqref{eqn:cor1} and \eqref{eqn:gc_ss4}, this amounts to finding the nearest node. So, the schedule that selects the nearest neighbor as the next node to be polled is the optimum schedule. We call this algorithm: \emph{Nearest Neighbor Next} or \emph{NNN}. For a desired value of network lifetime, the \emph{NNN} schedule will give the smallest value of $B(X_1, X_2, \ldots, X_N)$ and the smallest sum of the transmission times, so using the ``Channel Aware'' algorithm proposed in \cite{wowmom05}, we can prove that this schedule maximizes the network lifetime.

It should be noted that it is the special property of this problem, due to the correlation model and assumption above, that the schedule that minimizes $B(X_1, X_2, \ldots, X_N)$, also minimizes the sum of the transmission times, and subsequently maximizes the network lifetime. In general, it is not true that the schedule that minimizes $B(X_1, X_2, \ldots, X_N)$ also minimizes the sum of the transmission times.
\end{proof}

\emph{Part 2:} Without the assumption in \emph{Part 1}, here we consider the general problem. The following theorem proves that for the given spatial correlation model, the problem is $\cal{NP}$-hard.

\begin{pavikt}
The static scheduling problem in \eqref{eqn:gc_ss1} is $\cal{NP}$-hard for the spatial correlation model of \eqref{eqn:gc_ss4}.
\end{pavikt}
\begin{proof}
An arbitrary instance of ``Shortest Hamiltonian Path'' problem can be reduced to this problem by following the same sequence of steps as in the proof in \emph{Example 2}.
\end{proof}

\emph{Example 2:} Let us consider the spatial correlation model of \cite{Cristescu}, that is the one where the sensor data is modeled by Gaussian random field. Thus, we assume that an $N$ dimensional \emph{jointly Gaussian multivariate distribution} $f(\mathbf X)$ models the spatial data $\mathbf{X}$ of $N$ sensor nodes.
\begin{equation}
\label{eqn:cor2}
f(\mathbf X) = \frac{1}{(\sqrt{2\pi})^N \det(K)^{1/2}} e^{-\frac{1}{2}(\mathbf X - \mu)^TK^{-1}(\mathbf X - \mu)}
\end{equation}

where $K$ is the (positive definite) covariance matrix of $\mathbf X$, and $\mu$ the mean vector. The diagonal entries of $K$ are the variances $K_{ii} = \sigma_i^2$. The rest of $K_{ij}$ depend on the distance between the nodes $i$ and $j$: $K_{ij} = \sigma^2 \exp(-ad_{i,j}^2)$.

Without any loss of generality, here we use differential entropy instead of entropy, as we assume that the data at each sensor node is quantized with the same quantization step, and under such assumption, differential entropy differs from entropy by a constant.

\begin{pavikt}
\label{gc_ss_thrm}
The static scheduling problem in \eqref{eqn:gc_ss1} is $\cal{NP}$-hard for the spatial correlation model of \eqref{eqn:cor2} and energy function $f(h,x)$.
\end{pavikt}
\begin{proof}
Let us consider the decision version of this problem: does there exist a schedule $\pi$, for which the network achieves the lifetime $L$, with the following constraints?
\begin{eqnarray}
\label{eqn:gc_ss2}
f(h_{\pi(k)}, t_{\pi(k)})d_k & \le & \frac{E_k}{L}, \forall k = 1 \ldots N \\
t_{\pi(k)} & \ge & 0 \nonumber \\
\sum^N_{k=1} t_{\pi(k)} & \le & 1 \nonumber
\end{eqnarray}

We will prove the $\cal{NP}$-hardness of this problem as follows. We reduce an arbitrary instance of ``Shortest Hamiltonian Path Problem'' \cite{Rubin} over Euclidean and complete graph, which is well known to be $\cal{NP}$-complete, to some instance of the problem in \eqref{eqn:gc_ss2}. For the given instance of SHP problem, interpret the edge cost between nodes $i$ and $j$, as the spatial distance between the nodes $i$ and $j$ of our problem. So, as we visit a node $k$ in the SHP tour $\pi$, we can compute the conditional entropy $h_{\pi(k)}$ of that node using the knowledge of the model for spatial correlation among the sensor nodes as well as the history of the tour so far. With this computed conditional entropy value, using the first constraint of \eqref{eqn:gc_ss2}, we can compute the minimum time $t_{\pi(k)}$ that this node needs to transmit $h_{\pi(k)}$ bits of information to the base-station. So, for every schedule $\pi$, we can compute the sum of the minimum transmission times.

Let us consider an Euclidean, complete graph of $N=4$ nodes $[A, B, C, D]$ with symmetric edge costs. Let us consider two schedules $ABCD$ and $ABDC$. We have taken $N=4$ to only illustrate the main idea of the reduction, otherwise the approach is general enough to be applicable to the bigger networks. Let us assumes that the length of Hamiltonian path $d_{ABCD}$ for schedule $ABCD$ is less than $d_{ABDC}$ for the schedule $ABDC$. Now we are going to prove that the sum of transmission times for the schedule $ABCD$ is less than that for the schedule $ABDC$. For the spatial correlation model of interest \eqref{eqn:cor2}, for every schedule, we can compute the conditional entropies of every node based on all the nodes visited previously \cite{Cover}. For example, if the schedule is $[A, B, C]$ and $X_A, X_B, X_C$ denote their data samples, respectively, then:
\begin{eqnarray*}
h(X_A) & = & \frac{1}{2}\log(2\pi e), \mbox{ assuming } \sigma^2 = 1 \\
h(X_B/X_A) & = & h(X_A, X_B) - h(X_A) \\
           & = & \frac{1}{2}\log((2\pi e) \det(K_{AB})) \\
h(X_C/X_A, X_B) & = & h(X_A, X_B, X_C) - h(X_A, X_B) \\
                & = & \frac{1}{2}\log((2\pi e) \frac{\det(K_{ABC})}{\det(K_{AB})})
\end{eqnarray*}

where $K_{ABC}$ and $K_{AB}$, denote the covariance matrices of $X_A, X_B, X_C$ and $X_A, X_B$, respectively.

Assume that the transmission time $t_i$ of node $i \in [A, B, C, D]$ is exponentially dependent on the entropy $h_i$ of the node (this follows from the empirical results obtained after numerically solving the equation \eqref{energy_fn}). Let us denote the transmission times of the nodes A, B, C, and D under schedule $ABCD$ as $t_A, t_B, t_C,$ and $t_D$ respectively. Similarly, for the schedule $ABDC$, let the corresponding times be $t'_A, t'_B, t'_C,$ and $t'_D$ respectively. Note that $t_A = t'_A$ and $t_B= t'_B$. Now
\begin{eqnarray}
\label{eqn:gc_ss_times1}t_A + t_B + t_C + t_D & < & t'_A + t'_B + t'_C + t'_D \\
\label{eqn:gc_ss_times2}\mbox{if } t_C + t_D & < & t'_C + t'_D
\end{eqnarray}

After substituting the values of $t_C, t_D, t'_C$, and $t'_D$ and a little algebraic manipulation of the resulting expressions, we show that \eqref{eqn:gc_ss_times1} is true if (with $\sigma$ and $\alpha$ as defined in \cite{Cristescu}):
\begin{eqnarray}
& &e^{-2 \alpha d_{AC}^2}+e^{-2 \alpha d_{BC}^2} > e^{-2 \alpha d_{AD}^2}+e^{-2 \alpha d_{BD}^2} \\
&&\mbox{if } d_{AC}^2 + d_{BC}^2 < d_{AD}^2 + d_{BD}^2 \\
&&\mbox{if } d_{AB} + d_{BC} + d_{CD} < d_{AB} + d_{BD} + d_{CD}\\
&&\mbox{if } d_{ABCD} < d_{ABDC}
\end{eqnarray}

So, if a schedule has smaller Hamiltonian path length, then the corresponding sum of the transmission times will be smaller too. This implies that the solution of SHP gives the smallest value of the sum of the transmission times. So, for the schedule that gives shortest Hamiltonian path, we can compute the sum of the transmission times and if this sum is less than $1$, then we have at least one schedule that achieves the lifetime $L$.
\end{proof}

\subsection{Dynamic Scheduling}
\label{gc_ds}
In this section, we explore how network lifetime can be increased by employing multiple schedules. Instead of restricting the network to follow a single schedule, we allow the system to employ different schedules over time. There are $N!$ possible schedules to choose from. Let $h_{\pi(k)}$ be the number of information bits generated per slot by node $k$ under the schedule $\pi$, $1 \le k \le N$. Two or more schedules can collaborate by having the nodes use non-optimal transmit energies over two or more data-transmission slots to increase the lifetime of the network.

We have a total of $N!$ schedules. If only $m,\; 1 \le m \le N!$ schedules are going to cooperate, then there are $C(N!, m)$ possible combination of the schedules.  Let $\tau_{\pi_i}$ denote the number of time slots for which schedule $\pi_i, i \in \{1, \ldots, m\}$ is employed. The optimal network lifetime $L$ under dynamic scheduling is the solution to the following optimization problem
\begin{eqnarray}
\label{eqn:gc_ds}
& L = \max\limits_{\substack{m \\ 1 \le m \le N!}} \max\limits_{\substack{[\pi_1, \ldots, \pi_m] \\ \pi_1, \ldots, \pi_m \in \Pi}}\sum\limits_{i = 1}^m \tau_{\pi_i} &\\
&\mbox{s. t. } \sum\limits_{i=1}^m f(h_{\pi_i(k)}, t_{\pi_i(k)})d_k \tau_{\pi_i} \le E_k, \forall 1 \le k \le N & \nonumber \\
&\sum\limits_{k=1}^{N} t_{\pi_i(k)} = 1, \forall 1 \le i \le m & \nonumber
\end{eqnarray}

Specifically for $m = N!$, we have
\begin{eqnarray}
\label{eqn:gc_ds_all}
& L = \max \sum\limits_{i = 1}^{N!} \tau_{\pi_i} &\\
&\mbox{s. t. } \sum\limits_{i=1}^{N!} f(h_{\pi_i(k)}, t_{\pi_i(k)})d_k \tau_{\pi_i} \le E_k, \forall 1 \le k \le N & \nonumber \\
&\sum\limits_{k=1}^{N} t_{\pi_i(k)} = 1, \forall 1 \le i \le N! & \nonumber
\end{eqnarray}

Also note that if $m=1$, then the problem in \eqref{eqn:gc_ds} reduces to the static scheduling problem in \eqref{eqn:gc_ss1}. So, the computational complexity of this problem cannot be any less than that of the static scheduling problem, which is proven to be $\cal{NP}$-hard in Theorem \ref{gc_ss_thrm}. Here the question we are concerned with is that if the dynamic scheduling can indeed increase the network lifetime. In the following, we prove that even for the simplest case of the network of two nodes, it is indeed so.

\begin{pavikt}
\label{gc_ds_thrm}
For $N=2$, dynamic scheduling performs better than the optimal static scheduling.
\end{pavikt}
\begin{proof}
For the network of two nodes, let us consider two schedules $\pi_1$ where node $1$ is polled before node $2$ and $\pi_2$, where the nodes are polled otherwise. Now using our ``Channel Aware'' algorithm of the previous section, for a given polling schedule we can find the optimal allocation of the transmission times such that both the nodes spend same amount of energy, dying at the same time. Let for schedule $\pi_1$, this happens when the node 1 transmits for $t$ units of time and node 2 transmits for $1-t$ units of time. Similarly, for schedule $\pi_2$, let the corresponding times be $t'$ and $1-t'$. Let $h$ denote the entropy of first node polled in the schedule and $h_{1|2}$ denote the entropy of second node polled. So, for schedule $\pi_1$: $h_1 = h, h_2 = h_{2|1} = h_{1|2}$ and for schedule $\pi_2$: $h_1 = h_{1|2}, h_2 = h$. Given the optimality of $t$ and $t'$ for the schedules $\pi_1$ and $\pi_2$ respectively, we have for the energy consumptions of the nodes:
\begin{eqnarray}
\mbox{For schedule } \pi_1: f(h, t)d_1 = f(h_{1|2}, 1-t)d_2 \\
\mbox{For schedule } \pi_2: f(h_{1|2}, t')d_1 = f(h, 1-t')d_2
\end{eqnarray}

If we assume the schedule $\pi_1$ to be the optimum static schedule, then the following holds true:
\begin{equation*}
f(h, t)d_1 = f(h_{1|2}, 1-t)d_2 \le f(h_{1|2}, t')d_1 = f(h, 1-t')d_2
\end{equation*}

Let us consider the plot where the horizontal axis corresponds to the energy consumption $E_1$ of the node 1 and the vertical axis corresponds to the energy consumption $E_2$ of the node 2. In this plot, we draw the energy consumption curves for both the schedules $\pi_1$ and $\pi_2$, for different values of $t, 0 < t < 1$ and $t', 0 < t' < 1$ respectively. Given the form of the energy consumption curves, it is easy to verify that these two curves corresponding to two different schedules, are convex and will intersect at one and only one point.

Let us consider the ``equal energy line'' which passes through the pair of points $(f(h, t)d_1, f(h_{1|2}, 1-t)d_2)$ and $(f(h_{1|2}, t')d_1, f(h, 1-t')d_2)$, so the equation of this lines is
\begin{equation}
\label{line1}
E_1 = E_2
\end{equation}

Now let us also consider a line that passes through the point $(f(h, r)d_1, f(h_{1|2}, 1-r)d_2)$ on the curve corresponding to the schedule $\pi_1$ with $0 < r < 1$, and the point $(f(h_{1|2}, s)d_1, f(h, 1-s)d_2)$ with $0 < s < 1$ on the curve for schedule $\pi_2$. The equation for such a line is
\begin{eqnarray}
\label{line2}
E_2 = f(h_{1|2}, 1-r)d_2 + m(r, s)(E_1 - f(h, r)d_1 \\
\mbox{where } m(r, s) = \frac{f(h, 1-s)d_2 - f(h_{1|2}, 1-r)d_2}{f(h_{1|2}, s)d_1 - f(h, r)d_1} \nonumber
\end{eqnarray}

Now let us consider the point of intersection of these two lines. At the point of intersection, we have:
\begin{eqnarray}
\label{intersection}
E_1\!\!\! & = &\!\!\! E_2 \\
& = &\!\!\! \frac{f(h_{1|2}, 1-r)d_2 - m(r, s)f(h, r)d_1}{1 - m(r, s)} \nonumber \\
& = &\!\!\! \frac{f(h_{1|2}, 1-r)f(h_{1|2}, s)d_2 - f(h, 1-s)f(h, r)d_2}{f(h_{1|2}, s) - f(h, r)} \nonumber
\end{eqnarray}

Now, if we want to prove that with the dynamic scheduling we can perform better than the static scheduling, then we must prove that there exists at least one pair of values $(r, s)$, for which the following holds
\begin{eqnarray}
\label{ds_optimality}
E_1(r,s) < f(h, t)d_1 \\
E_2(r,s) < f(h_{1|2}, 1-t)d_2
\end{eqnarray}

Substituting the expressions of $E_1(r,s)$ and $E_2(r,s)$ from \eqref{intersection} in \eqref{ds_optimality}, and using the properties of the energy consumption function, we prove that the dynamic scheduling performs better than static scheduling for all $(r,s)$ such that
\begin{eqnarray}
r < t \\
\frac{ht - (h - h_{1|2})}{h_{1|2}} < s \nonumber
\end{eqnarray}
\end{proof}

This result implies that two schedule can cooperate to give longer network lifetime compared to optimum static schedule. Following subsection, generalizes this result.

\subsection{Geometrical Interpretation}
\label{geo_inter}
Let us consider the scenario where we have $N$ sensor nodes to poll and this polling can be done in $N!$ ways. Let us consider an $N$ dimensional Euclidean space, where an axis corresponds to the energy consumption of a node. Given that the energy consumption can only assume positive real values, we are only concerned with the first orthant of this $N$-dimensional space. For any given schedule, as the transmission time allocation to the different nodes changes, the corresponding energy consumption of the nodes changes. So, the point defining the energy consumption of the nodes in this $N$ dimensional space describes an $N$-dimensional convex hypersurface. For $N!$ possible schedules, we have $N!$ such hypersurfaces. The computational complexity of the problem of finding the optimum static and dynamic schedules depends on the properties of the intersections of these hypersurfaces. Also, note that the general shape of these surfaces is determined by the energy consumption function and the model of the spatial correlation in the sensor data.

The optimal static schedule is the one whose energy hypersurface is intersected by the ``equal energy line'' closest to the origin. Further, the dynamic scheduling helps us achieve all the points on the convex hull $\cal C$ of $N!$ convex hypersurfaces, as those are all the points achievable with the cooperation of any number of schedules. It is obvious that the network lifetime cannot be increased by the cooperation of those schedules and their transmission time allocations, which give any point in the interior of this convex-hull, as there is always a point on the surface of the convex-hull that is closer to the origin and gives better network lifetime. So, when two schedules cooperate, then the optimum transmission time allocation is the one that gives the line connecting the two points on the surface of the convex hull. Then the optimum network lifetime is achieved by some point on that line, specifically by the point on this line that is closest to the origin. Similarly, when three schedules cooperate, then the respective optimum transmission times allocation for the those three schedules is the one that gives the plane connecting the energy consumption points corresponding to the three schedules, on the surface of the convex hull and then the optimum network lifetime is achieved by some point on this plane.

Formally, if $\pi_1, \ldots, \pi_m, \pi_i \in \Pi$ are the $m$ cooperating schedules, then the plane defined by the $m$ points on the $m$ corresponding hypersurfaces, that is closest to the origin, must belong to the convex hull $\cal C$. For these $m$ schedules, the optimal lifetime of the network is obtained by that point on this plane that is closest to the origin. This is point is obtained as the solution to the following optimization problem:
\begin{eqnarray}
\label{geoeqn1}
\argmin_{r_1, \ldots, r_m} & & \sum_{i=1}^N (\sum_{j=1}^m r_jf(h_{\pi_j(i)}, t_{\pi_j(i)})d_i)^2 \\
\mbox{s. t. } & & 0 < r_j < 1, \forall 1 \le j \le m \nonumber \\
& & \sum_{j = 1}^{m} r_j = 1 \nonumber
\end{eqnarray}

With these optimum values of the parameters $\{r_j\}_{j=1}^m$, we solve for the network lifetime as follows:
\begin{eqnarray}
\label{geoeqn2}
L_{\pi_1\ldots\pi_m} & = & \!\!\!\!\!\! \min_{i \in \{1, \ldots, N\}} \frac{E_i}{\sum_{j = 1}^{m}r_jf(h_{\pi_j(i)}, t_{\pi_j(i)})d_i} \\
\mbox{s. t. } & & 0 < r_j < 1, \forall 1 \le j \le m \nonumber \\
& & \sum_{j = 1}^{m} r_j = 1 \nonumber
\end{eqnarray}

The optimum network lifetime $L_m$ for the set of $m$ schedules, is obtained by solving above set of equations for all $C(N!,m)$ possible combinations of $m$ schedules, that is $L_m$ is obtained by:
\begin{equation}
\label{geoeqn3}
L_m = \max\limits_{\substack{[\pi_1, \ldots, \pi_m] \\ \pi_1, \ldots, \pi_m \in \Pi}} L_{\pi_1\ldots\pi_m}
\end{equation}

Further, to obtain the optimum value of network lifetime over all possible combinations of the schedules is obtained by:
\begin{equation}
\label{geoeqn4}
L = \max_{1 \le m \le N!} L_m
\end{equation}

It should be noted that the equations \eqref{geoeqn1}-\eqref{geoeqn4}, essentially solve \eqref{eqn:gc_ds}. However, this alternative formulation of the problem in \eqref{eqn:gc_ds}, helps us to prove following two important theorems:

\begin{pavikt}
\label{geothrm1}
The optimum network lifetime for $m$ schedule cooperation is no worse than the optimum network lifetime for $m-1$ schedule cooperation. That is $L_m \ge L_{m-1}$.
\end{pavikt}
\begin{proof}
Omitted for brevity.
\end{proof}

\begin{pavikt}
\label{geothrm2}
When $m^*=m^*(N) < m \le N!$, then the cooperation among $m$ or more schedules does not improve the network lifetime anymore.
\end{pavikt}
\begin{proof}
Omitted for brevity.
\end{proof}

\section{Small Rate Region Approximation}
\label{srra}
In this section, we assume that transmission rate is linearly proportional to signal power. This assumption is motivated by Shannon's AWGN capacity formula which is approximately linear for low data rates. The energy expended by a node to transmit $H$ units of information is given by $H\times d$, where $d$ is the suitably normalized path loss factor between the node and the base station. The linear rate assumption implies, as shown below, that transmit energy is independent of transmission time. Hence, the optimal time allocation problem is trivial and we only need to find the optimal scheduling order.

For the small data rates, the energy consumption function for the $k^{\textrm{th}}$ node under schedule $\pi$ reduces to $f(h_{\pi(k)}, t_{\pi(k)}) d_k = h_{\pi(k)} d_k$. For example, by inverting Shannon's AWGN channel capacity formula, we get the following as the energy consumption function \cite{Prabhakar}:
\begin{equation*}
f(h_{\pi(k)}, t_{\pi(k)}) d_k = t_{\pi(k)}(2^{\frac{h_{\pi(k)}}{t_{\pi(k)}}} - 1) d_k
\end{equation*}
For the small data rates, this gives for some constant $c \ge 0$
\begin{eqnarray*}
f(h_{\pi(k)}, t_{\pi(k)}) d_k & \approx & t_{\pi(k)}(1 + c\frac{h_{\pi(k)}}{t_{\pi(k)}} - 1) d_k \\
                    & = & ch_{\pi(k)} d_k
\end{eqnarray*}

\subsection{Static Scheduling}
\label{srra_ss}
Under the ``small rate region approximation'', the static scheduling problem in \eqref{eqn:gc_ss} reduces to:
\begin{equation}
\label{eqn:srra_ss}
\max_{\pi \in \Pi} \min_{1 \le k \le N} \frac{E_{\pi(k)}}{h_{\pi(k)}d_{\pi(k)}}
\end{equation}
The objective function represents the lifetime of node $\pi(k)$ under the given static schedule $\pi$. In \cite{wowmom05}, we describe a greedy static scheduling strategy, Minimum Cost Next (MCN), and prove its optimality.

The MCN schedule not only maximizes the minimum lifetime, but also maximizes all lifetimes from $2^{nd}$ minimum lifetime to $N^{\textrm{th}}$ minimum lifetime. This is desirable in the situations, where the network has to continue to operate even when one or more nodes die out. Also, note that the MCN solution is Pareto-optimal. Given an MCN schedule, no other schedule can help increase any node's lifetime without decreasing some other node's lifetime.

\subsection{Dynamic Scheduling}
\label{srra_ds}
In this section, we explore how network lifetime can be increased by employing multiple schedules under the small data rate approximation. Under this assumption, as the general static scheduling problem in \eqref{eqn:gc_ss} reduced to \eqref{eqn:srra_ss}, the general dynamic scheduling problem in \eqref{eqn:gc_ds_all} reduces to the following optimization problem that gives the optimal lifetime $L$
\begin{eqnarray}
\label{eqn:srra_ds}
& L = \max \sum\limits_{\pi \in \Pi} \tau_{\pi} & \\
& \mbox{ s.t. }\sum\limits_{\pi \in \Pi} h_{\pi(k)} d_k \tau_{\pi} \le E_k,& \forall 1 \le k \le N \nonumber
\end{eqnarray}
$\tau_{\pi}$ is the number of slots for which the schedule $\pi$ is used. Once more, the constraints ensure that the time assignment is feasible for each node with respect to its energy capability. Also as in \ref{gc_ds}, \eqref{eqn:srra_ds} can be treated as a linear program. A dynamic schedule, $\tau$, is given by the set $\{ \tau_{\pi} \}$.

Given $N!$ variables, in general, there seems to be no easy way to solve \eqref{eqn:srra_ds} to compute the optimal $\tau_{\pi}$ values. However, in \cite{wowmom05}, we exploit the special nature of our problem to propose an algorithm, which we refer to as Lifetime Optimal Clustering ALgorithm (LOCAL), and prove its optimality.

\subsection{Geometric Interpretation}
\label{geo_inter_srra}
In this section we discuss both the static and dynamic scheduling in ``small rate region approximation'' in the spirit of discussion of the nature of the solutions in \ref{geo_inter}. Under this approximation, the energy consumption is independent of the transmission time, so the flexibility to change the energy consumption by varying the transmission time is not available. So, in the $N$-dimensional space, where each axis corresponds to the energy consumption of a node, for every schedule, we get a point in this space, rather than a hypersurface. For $N!$ possible polling schedules, we get $N!$ points.

The optimal static schedule under this approximation is given the MCN algorithm. Given the nature of the problem in \eqref{eqn:srra_ss}, this corresponds to that schedule which gives a point closest to the ``equal energy line''. This is not difficult to see if one notes that optimal static schedule attempts to `equalize' the lifetimes/energy consumption of the nodes. However, as noted above, the flexibility of varying the transmission times of the nodes to `equalize' their lifetimes is not available under this approximation, so the optimum static schedule does the best in providing a point closest to this ``equal energy line'', if not at the line itself. Note that the point corresponding to the optimum static schedule may not be the closest to the origin.

Similarly, the optimal dynamic scheduling under ``small rate region approximation'' corresponds to finding those schedules, the lines, planes, or the hyperplanes connecting which contain the point closest to the ``equal energy line''. When more than one such points are possible, then one that is closest to the origin is taken to be the point of operation of the network.

\section{Conclusions and Future Work}
In this paper, we have considered the problem of maximizing the lifetime of a data gathering wireless network. Our contribution differs from previous research in two respects. Firstly, we proposed a combined source-channel coding framework to mitigate the energy cost of radio transmission. Secondly, we have explicitly maximized network lifetime as opposed to other objective functions such as cumulative energy cost. To the best of our knowledge, both these aspects have not been explored in the context of sensor networks previously. In our system model, nodes communicate directly to a base station in a time division multiplexed manner. With our notion of instantaneous decoding, we show that the network lifetime maximization problem reduces to finding an optimal scheduling strategy (polling order and transmission time allocation).

We considered the general channel problem and proved that there the optimal static and dynamic scheduling problems are $\cal{NP}$-hard and the optimal dynamic scheduling strategy indeed does better than the optimal static scheduling strategy. Then we considered the scenario where the energy consumption is independent of transmission time. For both, the general channel problem and its approximated version, we provided the geometric interpretation of the optimal solutions and the solution search processes.

This paper assumed that source and channel coding is optimal, quantization is perfect and that a continuum of power levels can be employed. Network lifetime obtained under these assumptions is an upper limit to practically achievable performance. It would be useful to consider the network lifetime problem with more realistic constraints. Finally, the system model has to be generalized to the multi-hop case.

\end{document}